# Structures in the nuclear electric-dipole spectrum[*]

P. Papakonstantinou

*Institute for Basic Science, Rare Isotope Science Project,,
Daejeon 34047, South Korea*
[†]*E-mail: ppapakon@ibs.re.kr*

I undertake an anatomy of the nuclear electric-dipole spectrum and discuss selected observations. First, using data and calculations on $^{208}$Pb as a representative case, I identify five conspicuous structures in the spectra of heavy nuclei: (1) the giant dipole resonance (GDR), (2) the dipole compression mode, (3) the isoscalar low-energy dipole mode (IS-LED), (4) the IS mode in the region of the GDR, and (5) a concentration of "pygmy" dipole strength (PDS) between the latter two. Next, compiling the data from Texas A&M I show that the energy of (4) follows roughly a $A^{-1/6}$ dependence on the mass number. Finally, I summarize recent theoretical studies on the PDS in exotic nuclei, which suggest a strong influence of shell effects and loose binding.

*Keywords*: giant resonances; pygmy resonances; heavy nuclei; exotic nuclei; electric-dipole spectrum

## 1. Introduction

When nuclei interact with photons, electrons, or hadrons, several resonance-like structures are observed in the electric-dipole channel, besides the well-known giant dipole resonance (GDR). Comparisons with theoretical models help us understand the structures' physical origins, make connections with the nuclear equation of state, or improve the modelling of exotic r-proccess nuclei where threshold transitions may enhance the capture rates. In this contribution I undertake an anatomy of the nuclear electric-dipole spectrum and discuss selected open issues. First, using the many data and calculations on $^{208}$Pb as a representative case, I identify five structures in the E1 spectra of heavy nuclei: (1) the giant dipole resonance (GDR), (2) the dipole compression mode, (3) the isoscalar low-energy dipole mode (IS-LED), (4) the IS mode in the region of the GDR, and (5) a concentration of "pygmy" dipole strength (PDS) between the latter two. Next, compiling the data from Texas A&M I suggest that the energy

---

[*] Presented at the VIII International Symposium on EXOtic Nuclei (EXON 2016), Kazan, Russia,

of (4) follows roughly a $A^{-1/6}$ dependence on the mass number. Finally, I focus on the PDS and summarize recent theoretical studies on exotic nuclei, which show the strong influence on PDS of shell effects and lose binding.

## 2. Richness of the dipole spectrum – surveying $^{208}$Pb

The upper panel in Fig. 1 shows the photoresponse of $^{208}$Pb below and above the particle-emission threshold ($E_{th}$) as measured in various experiments or evaluated and available in the EXFOR database. The major structure at 13-14 MeV, labelled (1), is of course the giant dipole resonance (GDR), classically interpreted as an out-of-phase oscillation of the proton and neutron fluids [1]. Near or below $E_{th}$ there are several peaks which may or may not belong to the tail of the GDR, and which I refer to as pygmy dipole strength (PDS) following common practice.

The shown spectrum contains a wealth of information. The summed transition strength, weighted with the excitation energy, provides the total photoabsorption cross section. A comparison with the classical Thomas-Reiche-Kuhn sum rule reveals the non-local character of nuclear interactions [1]. The inverse energy-weighted sum, on the other hand, yields the dipole polarizability $a_D$ of the nucleus. Theoretical studies have linked $a_D$ with the thickness of the neutron skin and the slope of the symmetry energy [2,3], which affects the structure of neutron stars. (A complete measurement of the $^{208}$Pb spectrum via polarized proton scattering has provided us with a datum for the polarizability of this nucleus [4].) Finally, transitions near and below $E_{th}$ in the case of r-process nuclei can greatly affect capture rates and therefore the nucleosynthesis yields [5]. This seemingly simple spectrum is therefore rich in consequences. But how well do we understand its structure and origins?

We may obtain many clues by looking at the dipole spectrum through the lens of different probes, in particular isoscalar probes. Alpha [6] and oxygen [7] scattering have revealed the isoscalar response of $^{208}$Pb represented in the lower panel of Fig. 1 by the energy weighted transition sthrength. The bimodal structure occuring above threshold has been observed in several nuclei [8]. Theoretical studies agree that the major structure beyond 20 MeV, labelled (2), corresponds to a dipole compression mode [1], whereby a sound wave propagates back and forth inside the nucleus. The origin of the peak in the region of the GDR (4) has been attributed to toroidal modes [9] and other kinetic effects [10]. We shall return to this feature.

As observed in Ref. [7] in the case of $^{208}$Pb, but also in other heavy nuclei studied elsewhere, the PDS region presents a kind of bimodal structure too: Of all the participating transitions, only the lowest-energy ones respond strongly to the isoscalar probe. Thus two different "pygmy" regimes are revealed, one strongly isoscalar, and one isovector-only (5).

The isoscalar part (3), to which I referred to as IS-LED above, is quite a universal resonance in stable nuclei and possibly many unstable nuclei, appearing typically at 6-7MeV and exhausting several percentage points of the isoscalar EWSR – but less than one single-particle unit in the isovector channel. Linear-response theory reproduces its properties as a surface, tidal dipole mode, except that it systematically overestimates its energy (cf. lowest peak in Fig. 3). The interested reader may consult Ref. [11] for a brief survey, a compilation of data and bibliography. The isovector part is a complex problem to which we shall return shortly.

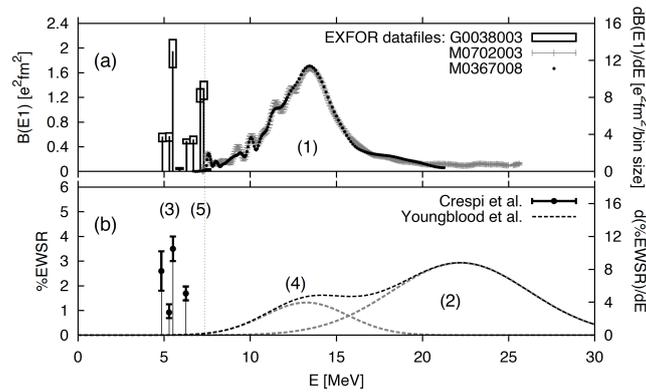

Fig. 1. Electric dipole spectrum of 208Pb. (a) Photoresponse $B(E1)\uparrow$ ; data were taken from the EXFOR database as indicated. (b) Isoscalar energy-weighted contributions. Shown are the major peaks identified in Ref. [Crespi] below the particle threshold and the Gaussian parameterization of the Ref. [Youngblood] data above threshold. Both (a),(b): The left (right) axis corresponds to the data below (above) the particle threshold, which is marked by a vertical line. Numbers mark the regimes of interest: (1) GDR, (2) compression mode, (3) IS-LED, (4) other isoscalar mode, (5) PDS .

## 3. The isoscalar mode in the GDR region

The isoscalar structure in the region of the GDR has been observed in many nuclei via alpha scattering at Texas A&M [6,8]. Early theoretical predictions did not distiguish between this and the IS-LED (3), though toroidal motion has remained a candidate as a generating mechanism [12] since it was first proposed [9]. RPA calculations in a harmonic-oscillator basis (HO) and coordinate-space RPA calculations with proper boundary conditions in the continuum (CRPA) are shown in Fig. 2. Both implementations predict IS strength in that energy region, in general agreement with other calculations. CRPA generates somewhat more suprathreshold strength. The isospin character of the transitions is found largely mixed, which means that they contribute to the GDR strength.

Fig. 3 shows a compilation of existing data. A comparison between Fig. 3(a) (plot of $A^{1/3} E$) and 3(b) ($A^{1/6} E$) suggests that the energy E of the unknown mode follows rather a $A^{-1/6}$ dependence on the mass number, possibly signifying a surface phenomenon. This is an open theoretical issue.

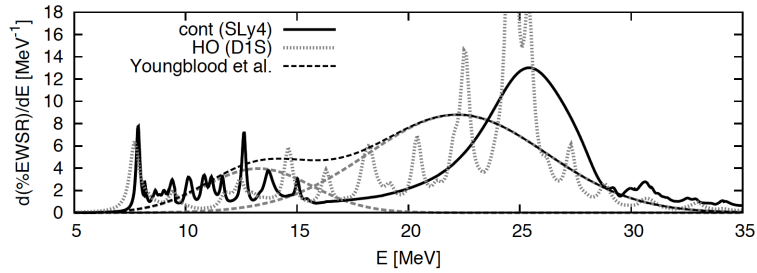

Fig. 2. For $^{208}$Pb: Distribution of the isoscalar energy-weighted sum as extracted from the data above particle threshold (two-Gaussian parameterization) [6] and as calculated within RPA in a harmonic-oscillator basis with the Gogny D1S interaction or in coordinate space with continuum effects with the Skyrme SLy4 interaction.

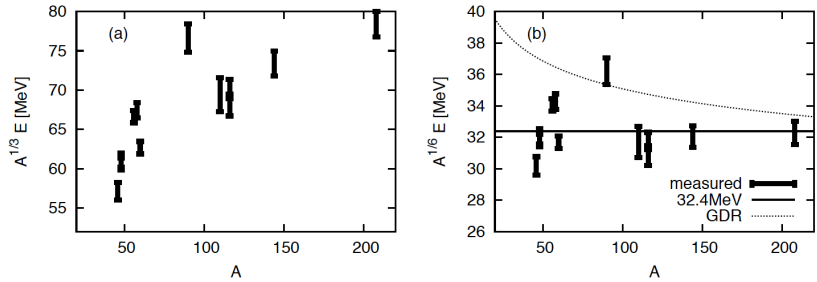

Fig. 3. The energy of the IS dipole mode in the GDR region multiplied by (a) $A^{1/3}$ and (b) $A^{1/6}$. On average $E \sim A^{-1/6} 32.4$ MeV. The data are from Refs. [6,8]. The empirical GDR energy, $A^{1/6} E \sim (31.2 A^{-1/6} + 20.6)$ MeV, is shown too.

## 4. Pygmy dipole strength and exotic nuclei

Let us now return to the isovector part of the PDS. Knowledge of its generating mechanism is important for establishing physical connections of PDS to other observables or parameters, such as the symmetry energy and its density dependence. At the same time, it is important to identify factors which might contaminate correlations between PDS and other observables.

A candidate mechanism for generating PDS is single-particle strength (E~1ℏω), as was demonstrated early on [13]. Other non-relativistic models have found that the PDS is centered around E~1ℏω [14]. Another popular mechanism, a neutron-skin oscillation, is less favored, because it would respond strongly to the IS field. In stable nuclei such a mode might contribute at energies near the neutron-emission threshold [15] or above it. It may generate the pygmy resonances observed in exotic Sn isotopes [16] and $^{68}$Ni [17,18].

In Ref [19] it was shown that, along the Ni isotopic chain and according to RPA calculations, neutron excess influences the PDS or the polarizability and the neutron skin thickness in different ways or degrees, apparently due to the different manifestations of shell filling. Correlations along isotopic chains are thus not straightforward, even though correlations for a single nucleus can be theoretically identified (their model dependence notwithstanding). Measurements of PDS on either side of a shell closure such as N=50 (Ni) or N=82 (Sn) and in the same energy interval are advocated, so as to test theoretical predictions on the strong influence of shell filling. Beyond the shell closures, correlations between the neutron-skin thickness and the PDS are predicted to be cleaner [20].

The relevance of separation energies has been put forth, e.g., in [21]. Loose binding can be a strong factor, compared with the (absolute) isospin asymmetry, in generating PDS and even collective skin modes. This was shown in Ref. [22] via a comparison between the stable Ca isotopes and their (mostly unstable) N=20 mirror isotones. Ca isotopes are excited weakly in terms of B(E1) below 9 MeV and do not show any evidence of skin modes. Proton-rich N=20 isotones, on the other hand, are theoretically expected to be much more strongly excited and some of them (e.g., $^{46}$Fe) to develop proton-skin modes [22]. The proton $f_{7/2}$ shell can be so loosely bound and spatially extended, that an extended proton density distribution can indeed develop.

## 5. Summary

I distinguish five salient structures in the electric-dipole spectra of heavy nuclei and discuss related open issues. The physical origin of the IS structure in the GDR region remains of interest. The pygmy dipole strength is strongly affected by shell structure and loose binding in exotic systems.

This was a selective presentation ommitting, e.g., soft modes in light halo nuclei and alpha-clustering effects and, on the theoretical side, the effects of phonon coupling.

**Acknowledgments**

I am thankful to all my collaborators for valued interactions in this line of work. This work and the presentation at EXON 2016 were supported by the Rare Isotope Science Project of the Institute for Basic Science funded by the Ministry of Science, ICT and Future Planning and the National Research Foundation (NRF) of Korea (2013M7A1A1075764).